\documentclass{PoS}

\usepackage{wrapfig}
\usepackage[numbers]{natbib}

\title{Neutron Star Merger Afterglows: Population Prospects for the Gravitational Wave Era}

\ShortTitle{Neutron Star Merger Afterglows in the GW Era}

\author{\speaker{Rapha\"el Duque}, Robert Mochkovitch, Fr\'ed\'eric Daigne\\
        Sorbonne Universit\'e, CNRS, UMR 7095, Institut d'Astrophysique de Paris, 98 bis boulevard Arago, 75014 Paris, France\\
        E-mail: \email{duque@iap.fr}}

\abstract{Following the historical observation of GW170817 and its electromagnetic follow-up, new neutron star merger afterglows are expected to be observed as counterparts to gravitational wave signals during the next science runs of the gravitational interferometer network. The diversity of the observed population of afterglows of these future events is subject to various factors, which are (i) intrinsic, such as the energy of the ejecta, (ii) environmental, such as the ambient medium density or (iii) observational, such as the viewing angle and distance of the source. Through prescribing a population of mergers and modelling their afterglows, we study the diversity of those events to be observed jointly in gravitational waves and electromagnetic bands. In the future, observables of detected events such as viewing angle, distance, afterglow peak flux or proper motion will form distributions which together with predictions from our study will provide insight on neutron star mergers and their environments.}

\FullConference{The New Era of Multi-Messenger Astrophysics - Asterics2019\\
		25 - 29 March, 2019\\
		Groningen, The Netherlands}

\begin{document}

\section{Introduction}

On August 17\textsuperscript{th} 2017, historical observations of the merger of a binary neutron star (BNS) were made. These were triggered by the detection by the interferometers of the LIGO-Virgo Collaboration (LVC) of the gravitational wave (GW) signal from the inspiral phase of the merger \cite{23}. These GW were followed by three electromagnetic counterparts: (i) a short and hard gamma ray burst (GRB) \cite{52, 137}, associated to high-energy dissipation processes in an ultrarelativistic jet from the merger, (ii) a thermal emission \cite[e.g.][]{57, 38}, powered by the radioactive decay of heavy nuclei synthesized through the $r$-process in the neutron-rich merger ejecta, reddening and fainting on the scale of tens of days and known as a kilonova (KN), and (iii) a multi-wavelength (radio to X-ray) long-lived signal known as the afterglow (AG) \cite{8, 12, 135}, produced by the synchrotron emission of particles accelerated at the forward shock formed as the merger ejectas penetrate the circum-merger medium.

These observations were historical in many regards. First of all, they confirmed BNS mergers as the progenitors of short GRBs, as hypothesized early in the history of GRB science and indirectly confirmed since then. Secondly, they inaugurated the era of multi-messenger astronomy with GW, and exposed the latter as an invaluable tool to study BNS mergers and short GRBs. This single event has had enormous repercussions in numerous branches of physics and astrophysics. To state only the most noteworthy of these, this event allowed to furthermore test General Relativity, to constrain the equation of state of hyperdense matter, to make a stand-alone measurement of the Hubble constant and to find new evidence for the astrophysical site of $r$-process nucleosynthesis.

The KN and AG provided a wealth of information on this event: its sub-arcsecond localization, external medium density, the kinetic energy of the ultrarelativistic jet, the viewing angle, etc. A breakthrough observation was the very long base interferometry imaging of the remnant through the radio AG \cite{79, 110}, which confirmed the emergence of the jet and somewhat constrained its structure and our viewing conditions.

On April 1\textsuperscript{st} 2019 the LVC started the third GW observing run (O3). It will run for a year with the two LIGO and the Virgo interferometers all with enhanced sensitivities with respect to O2. We thus expect more GW from BNS mergers, and KN and AG counterparts. Given the 170817 event and our knowledge of BNS mergers from short GRB science, what AG and KN should we expect in this new O3 run? What will these future events' AGs look like? How will they help us to study the evolution and environment of BNSs? What will they teach us more on GRBs and their dissipation mechanisms?

We address these questions with a population model in the multi-messenger approach, as illustrated in \cite{Duque}.

\section{From the intrinsic population to multi-messenger events}
\subsection{Multi-messenger detection criteria}

We model the GW, radio AG and KN emissions of individual events using a synthesis of prior knowledge on GRB afterglows and observations from 170817. These three signals depend on the event's intrinsic parameters such as the jet's kinetic energy and opening angle, on external parameters such as the density of the circum-merger medium, and finally on our observing conditions such as the luminosity distance to the merger and the viewing angle (from our line-of-sight to the jet axis). For the AG, we suppose that the jet's core's contribution to the flux dominates that of an eventual lateral structure at the peak of the AG, when the signal is most likely to be observed. 

We prescribe initial distributions on all of these parameters thanks to prior knowledge from GRB science and observations of the 170817 event. These make up an intrinsic population of mergers which occur in the local Universe with a rate of $1540^{+3200}_{-1220}~{\rm Gpc}^{-3}{\rm yr}^{-1}$, as inferred by the LVC during the O1 and O2 runs \cite{23}. The GW, AG and KN of these events are not all detectable because of detector limitations. For the GW, this is described by the interferometer's horizon, for the AG by the radio array's limiting sensitivity (at 3 GHz in our case) and for the KN by the depth of the visible-IR follow-up searches. Applying these detection criteria for the detection of the GW, the AG and the KN allows us to determine those events detectable in these messengers. Whether these detectable events will be detected is another issue we will discuss below. Since the follow-up of these events is triggered by the GW, we select the events observable in GW \textit{and} through their AG or KN. These constitute a population of jointly detectable events which we will study here.

As the detectors improve, this population will change, and we may thus make prospects for the expected detectable population in different multi-messenger detector configurations.

\subsection{The kinetic energy distribution}

To illustrate our choice of a parameter distribution for the intrinsic population of mergers, the jet kinetic energy distribution is directly deduced from prior work on the luminosity function (LF) of short GRBs. This is a fundamental and uncertain quantity of GRB science and has been constrained from observations by various authors. For a given GRB, the kinetic energy available to the jet for the AG is the energy which remains after dissipation in gamma-rays. Starting from a short GRB luminosity function, and supposing a typical duration of $0.2~{\rm s}$ and gamma-ray efficiency of $20\%$, we obtain a distribution for the kinetic energy of the merger jets. Using the short GRB LF of \cite{104}, we obtain the distribution reported in \cite[][Fig. 8a]{Duque}. Along with that of the kinetic distribution, we specify distributions on the circum-merger medium density and the shock microphysics parameters. It is noted in \cite{Duque} that the selection by radio detection favors the higher-energy jets, and this is reflected in the kinetic energy distribution of the detectable population.

\section{Results}
\subsection{Joint event rates}
Using our population model, we can predict the rates of jointly detectable events. Assuming a limiting radio sensitivity at the level of the Very Large Array ($10~\mu$Jy) and a GW horizon at the O3 level ($\sim~250~{\rm Mpc}$), we predict that $20-40\%$ of GW-detected events should have a detectable radio AG. This depends on the population model, and the actual fraction after observing many events should be considered as an insight on the jet kinetic energy function, and thus on the short GRB LF. The one-year O3 run should produce $\sim~9$ BNS GW events \cite{54}, of which $2-4$ should have a detectable AG. These numbers increase with the GW horizon, and the design-level interferometers (horizon of $\sim~450~{\rm Mpc}$) coupled to the Square Kilometer Array of the late 2020s (sensitivity of $\sim~1~\mu{\rm Jy}$) should produce $\sim~20$~BNS GW triggers per year of operation \cite{54}, of which $2-6$ with a detectable radio AG. We note that in all of these detector configurations, the principal factor limiting joint detections is the radio detection.

There remains nonetheless the question of whether these will be detected. This is a pivotal question because the detection, follow-up and astrophysical treatment of the AG requires the pin-pointing of the source, and thus to search and detect the KN within the GW-inferred sky-map of the signal. We will come back to this in Sec.~\ref{KN}.

\subsection{On the viewing angle of future events}

\begin{wrapfigure}{r}{0.4\linewidth}
\includegraphics[width=1.\linewidth]{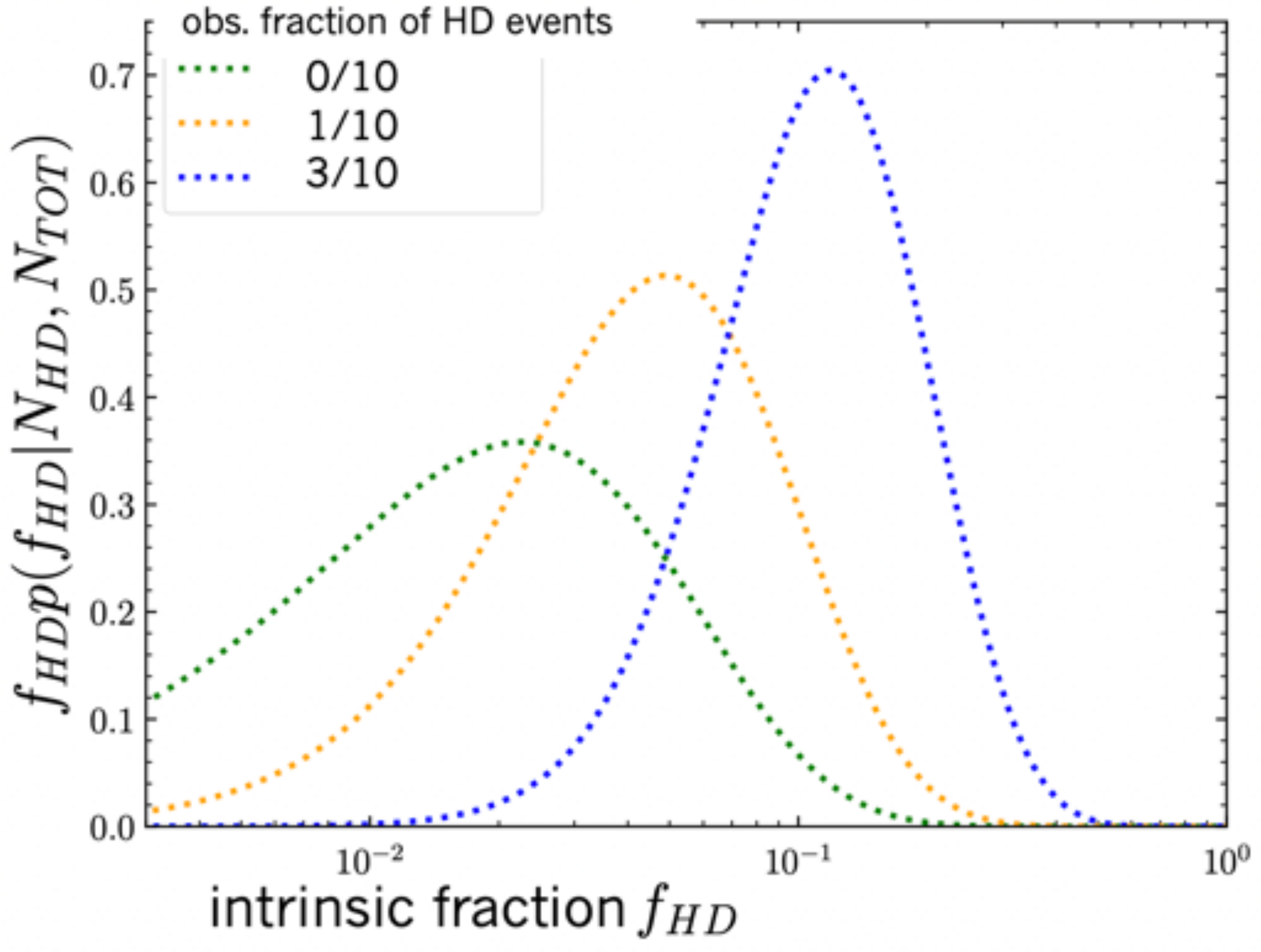}
\caption{Posterior probability distribution of the intrinsic fraction of high-density ($1~{\rm cm}^{-3}$) mergers obtained after having observed 0 (green), 1 (orange) or 3 (blue) of these among a total of 10 events, the others being at low density ($10^{-3}~{\rm cm}^{-3}$).}
\label{fpf}
\end{wrapfigure}

The radio detection strongly biases the observed population towards smaller viewing angles. Nonetheless, the configuration of the O3 run should provide events with a mean viewing angle over $20^\circ$, the value assumed by GRB170817A.

These events should show long increasing phases of their light-curves, thus allowing a detailed study of the jet structure. This could possibly give insight on the dissipation mechanisms at play in the lateral structure of the jet, which may have produced the soft tail observed in GRB170817A \cite{52} and found in some short GRBS by archival studies \cite{96}.

Also, we predict that $\lesssim 10\%$ of the joint events to come should be seen on-axis (our line-of-sight within the jet), likely resulting in \textit{classical} GRB counterparts. We note that our figure is consistent with rates of GW-GRB associations found by others with a different approach \cite{67}.

\subsection{Insight on fast-merging binaries}

Another bright insight of this population study is on the evolution of BNSs. Those which got through an efficient common envelope phase or acquire a large eccentricity after the second supernova merge after only a short delay. Many studies have found evidence for a excess of such binaries in the population \cite[e.g.][]{158, 155}. Such short-delay binaries are likely to merge in higher density environments, because they are allowed only a short migration from their birth site to their merger loci.

Since the AG peak flux depends strongly on the medium density ($F_p \propto n^{4/5}$), mergers in high densities produce brighter AGs and a population of short-delay (i.e. high-density) mergers should be over-represented in the observed population with respect to their intrinsic fraction. Thus, tight constraints on this population of fast-merging binaries are obtained with only a small number of joint events. This is illustrated in Fig.~\ref{fpf}, which shows that a 1-$\sigma$ uncertainty of $\sim~0.3~{\rm dex}$ can be obtained on the intrinsic fraction of high-density mergers after observing only 3 high-density mergers among 10 joint events. The observation of such AGs would be an independent proof of the existence of short-delay binaries and their study through this method a new insight on this particular evolution channel of BNSs.

\subsection{A word on kilonovae and on detecting the detectable}
\label{KN}

The observations of the KN of 170817 \cite{57} and prior theoretical considerations led to the idea that the outflows from the merger producing the KN emission were of two kinds, with different origins, physical characteristics and visible-IR signatures. The first is equatorial, composed of tidal ejections from the late inspiral phase and of dynamical ejections produced upon merger, and rather electron-poor. Its opacity is large, and has a red, slowly-varying emission. The other is rather polar, and its origin is not clear. Its electron fraction is high, leading to a low lanthanide enrichment, a lower opacity and thus a faster-evolving blue color.

\begin{wrapfigure}{l}{0.4\linewidth}
\begin{center}
\includegraphics[width=1.\linewidth]{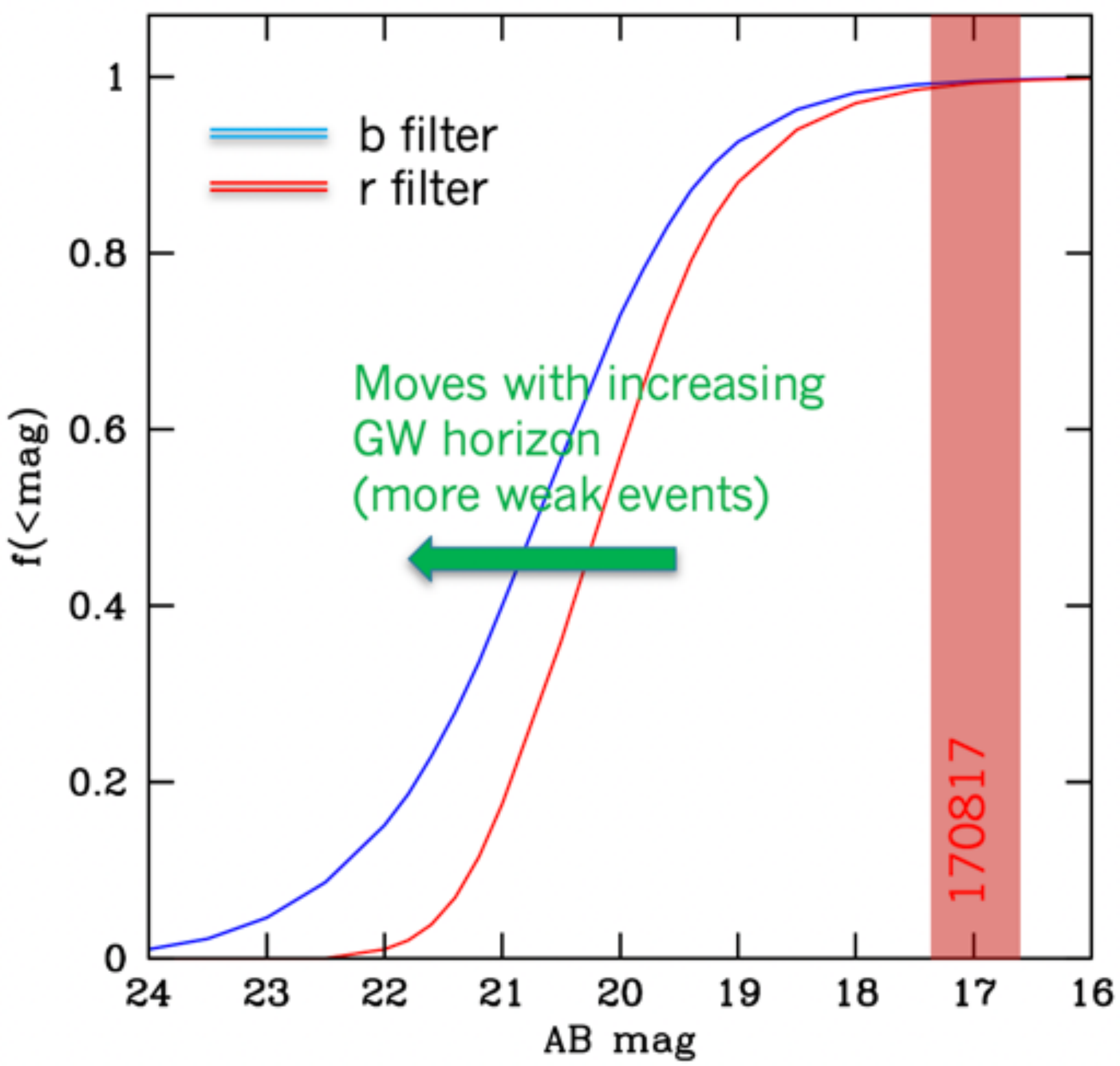}
\end{center}
\caption{Cumulative distribution of the expected magnitudes of the KN signals counterparts to GW triggers in the O3 run, in the $b$ (blue) and $r$ (red) band. The discovery magnitude of KN170817 is indicated for comparison.}
\label{mag}
\end{wrapfigure}

In consequence, the visible-IR signal from the KN also depends on the viewing angle to the system. Adopting a model for this angular dependence of the KN signal \cite{142}, one may determine the distribution of the magnitudes to expect for the KN as counterparts to GW in the O3 run and beyond. This is illustrated in Fig.~\ref{mag} for the O3 configuration in the $b$ and $r$ bands. Assuming a limiting follow-up magnitude of $\sim~21$, we expect the KN to remain in reach, up to design-level GW horizons.

Nonetheless, the detection of the KN may reveal challenging for two reasons. The first is the typical volume to explore during the O3 run to find the KN. In the case of GW170817, the event's proximity ($\sim~40~{\rm Mpc}$) allowed for a small GW-inferred localization map ($\sim~30~{\rm deg}^2$), and the prompt localization of the KN at a magnitude of $\sim~17$ in NGC4993. During O3, assuming a constant relative uncertainty $\Delta D / D \sim 25\%$ on the GW-inferred distance and the median localization prospects \cite{54}, the volume (and thus the number of galaxies) to explore in search of the KN will be at least $100$ times larger than in the case of GW170817. This can be a severe limitation to the pin-pointing of the event, and thus to the AG detection. This localization limitation should be hindered when more interferometers join the the network. The first should be the Japanese KAGRA, in the early 2020s.

The second challenging aspect is the contrast of the KN signal with respect to the host galaxy. The KN is a point source inside the extended emission of the host galaxy, and the distance to the object plays strongly against its detection.

\section{Conclusion}

In conclusion, though there should be difficulty in localizing the binary neutron star mergers, the effort is well worth it. Indeed, regardless of the evolution of gravitational wave and radio detectors, multi-wavelength afterglows and kilonovae will remain instrumental in the study of both the formation and the merging of binary neutron stars.

\newpage

\bibliographystyle{JHEP}
\bibliography{asterics2019}

\end{document}